\documentclass[a4paper]{article}

\usepackage{INTERSPEECH2022}

\usepackage{multirow} 
\usepackage{cite} 
\usepackage{subfigure} 
\usepackage{makecell} 
\usepackage{float} 
\usepackage{enumitem} 

\title{Environment Aware Text-to-Speech Synthesis}
\name{Daxin Tan, Guangyan Zhang, Tan Lee}

\address{Department of Electronic Engineering, The Chinese University of Hong Kong, Hong Kong}
\email{\{daxintan, gyzhang\}@link.cuhk.edu.hk, tanlee@ee.cuhk.edu.hk}

\begin{document}

\maketitle

\begin{abstract}
This study aims at designing an environment-aware text-to-speech (TTS) system that can generate speech to suit specific acoustic environments. It is also motivated by the desire to leverage massive data of speech audio from heterogeneous sources in TTS system development. The key idea is to model the acoustic environment in speech audio as a factor of data variability and incorporate it as a condition in the process of neural network based speech synthesis. Two embedding extractors are trained with two purposely constructed datasets for characterization and disentanglement of speaker and environment factors in speech. A neural network model is trained to generate speech from extracted speaker and environment embeddings. Objective and subjective evaluation results demonstrate that the proposed TTS system is able to effectively disentangle speaker and environment factors and synthesize speech audio that carries designated speaker characteristics and environment attribute. Audio samples are available online for demonstration \footnote{https://daxintan-cuhk.github.io/Environment-Aware-TTS/}.

\end{abstract}


\vspace{-0.5em}
\section{Introduction}
\vspace{-0.5em}
With the advancement of deep learning, neural network based text-to-speech synthesis (TTS) systems have attained great successes \cite{shen2018natural, yu2020durian,ren2020fastspeech}. They produce high-quality synthesized speech comparable to human speech in terms of intelligibility, naturalness and speaking style \cite{li2021controllable}. In most cases, TTS systems are trained with a large amount of clean speech with coherent audio quality. Naturally the training objective is to generate speech with similar acoustic characteristics to the speech for training.

Despite being widely accepted, this paradigm is found to pose various practical limitations that hinder the practical deployment of TTS systems. On one hand, collecting large amount of clean speech is generally difficult and costly, as professional arrangement of recording in well-controlled acoustic environment like soundproof studio is required. In contrast, realistic speech audios captured casually in daily-life environments, e.g., meeting room, play room, are much easier to access and accumulate. These speech signals are unexceptionally contaminated by background noise and reverberation. Being able to utilize speech data from heterogeneous sources is highly desirable for efficiently and maximally exploiting available data resources for TTS system development. 

On the other hand, it is an appealing function that a TTS system is able to synthesize speech to fit a designated acoustic environment. An example scenario is when a user needs to insert a speech segment into or substitute certain part of a recorded audio in order to change its speech content. The inserted or substituting segment can be synthesized from a given TTS system \cite{tan2021editspeech, jin2017voco}. The newly synthesized speech is required to carry the acoustic environment characteristics that are consistent with the original recording. In the application of voice cloning, it is common that only reverberant and/or noisy speech are available for the target speaker, while the goal is to generate clean speech of this speaker \cite{cong2020data, zhang2021denoispeech}. If ``clean'' is considered as one of the acoustic environment, this task can be formulated as generating speech to fit an acoustic environment that is different from the original one.

In the present study, we propose a method to utilize diverse-environment data for system development and extend the usage of TTS system. The main idea is to model acoustic environment of speech recording as a factor of data variability and incorporate it as a condition into the speech generation process. Specifically, \textbf{environment} in this study refers only to the room \textbf{reverberation}, while \textbf{noise} will be discussed in the future. In reality, speech data from a specific speaker are usually recorded in a limited range of environments, the speaker factor and the environment factor tend to be highly correlated. In the extreme case, speech from each speaker is associated with one single environment. Concerning this issue, the disentanglement between speaker and environment factor of speech is also tackled in this work. 

Modelling of environment-related factor in speech audio was investigated in previous studies \cite{hsu2019disentangling,cong2020data,zhang2021denoispeech, giri2019attention, hsieh2020wavecrn, williamson2017time,su2020acoustic}. In \cite{hsu2019disentangling}, disentanglement of speaker and noise was achieved through data augmentation and factorization. In \cite{cong2020data}, domain adversarial training was applied. Both studies were focused on making good use of data sources and aimed to generate clean speech for target speakers when only noisy speech data are available. In \cite{williamson2017time,su2020acoustic,wang2020deep}, the main goal was to reduce reverberation in speech. The task can be regarded as signal conversion from reverberant environment to ``clean'' environment. In \cite{su2020acoustic}, the problem of acoustic matching, i.e., conversion among different indoor acoustic environments, was investigated. These conversion processes operate typically in the waveform-to-waveform manner. In our proposed approach, acoustic environment is considered as a factor of variation in neural TTS.

We design a TTS model that leverages both clean and reverberant speech data. The model is trained to synthesize speech for target speakers under designated acoustic environments. Two extractor modules trained with specially designed datasets are used to extract embeddings that represent speaker and environment respectively. The TTS module generates speech by conditioning on the two embeddings. With speaker and environment information provided by reference utterances, the synthesized speech is made to carry the desired speaker characteristics and environment attributes. 

\begin{figure}[h]
  \centering
  \includegraphics[width=\linewidth, trim=40 330 40 10]{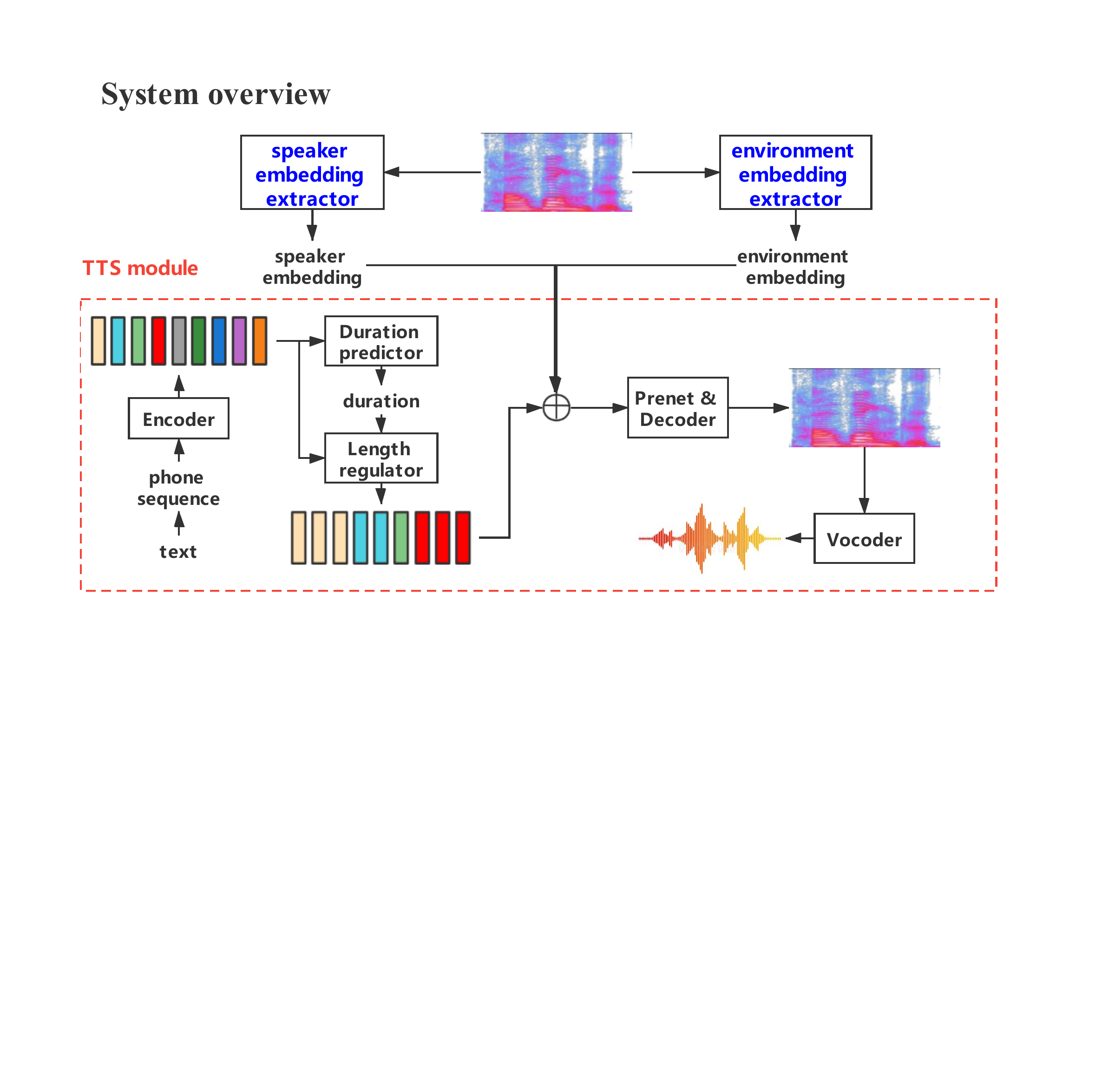}
  \caption{An overview of the proposed system.}
  \label{fig:system_overview}
  \vspace{-1em}
\end{figure}

\vspace{-1em}
\section{The proposed system}
An overview of our proposed system is shown in Figure \ref{fig:system_overview}. The system contains a speaker embedding extractor, an environment embedding extractor and a TTS module.

\vspace{-1em}
\subsection{Speaker embedding extractor}
\label{sec:speaker_embedding_extractor}
The speaker embedding extractor is developed based on a speaker verification (SV) system. The SV system is trained with the generalized end-to-end loss as proposed in \cite{wan2018generalized}. The objective is to make the embeddings extracted from utterances of the same speaker be close to each other, and enlarge the distance among the embeddings from different speakers. 

In each training batch, $s$ speakers are selected and $u$ utterances are sampled from each of the speakers. For each utterance, a speech segment of 80-frames length is cropped from random location of the utterance and is used as the input of extractor. The extractor model consists of LSTM layers followed by a linear layer. The output of extractor is L2-normalized to be the speaker embedding of this utterance. Let $E_s^{ij}$ denote the speaker embedding extracted from utterance $j$ of speaker $i$, where $1 \leq i \leq s$ and $1 \leq j \leq u$. For speaker $i$, the centroid of speaker embedding can be calculated in two different ways: (1) derived from all utterances of speaker $i$, denoted as $C_s^i$; (2) derived in a similar way to (1) but with utterance $j$ excluded, denoted as $C_s^{i,(-j)}$. The speaker embedding $E_s^{ij}$ is compared with the centroid of all speakers to construct the similarity matrix, which is then utilized to construct the softmax loss. The centroids, similarity matrix and loss function are defined as follows.

\noindent \textit{Centroid:}
\begin{center}
$C_s^{i}=\frac{1}{u} \sum\limits_{1\leq m\leq u} E_s^{im}, \quad 
C_s^{i,(-j)}=\frac{1}{u-1} \sum\limits_{\substack{1\leq m\leq u, m \ne j}} E_s^{im}$
\end{center}

\noindent \textit{Similarity matrix of speaker embeddings:}
$$ S_s^{ij,k}=\left\{
\begin{array}{rcl}
w*cos(E_s^{ij}, C_s^{i,(-j)})+b, & k=i\\
w*cos(E_s^{ij}, C_s^k)+b, & k\ne i 
\end{array}
\right.$$

\begin{figure}[h]
  \centering
  \includegraphics[width=\linewidth, trim=0 30 0 0]{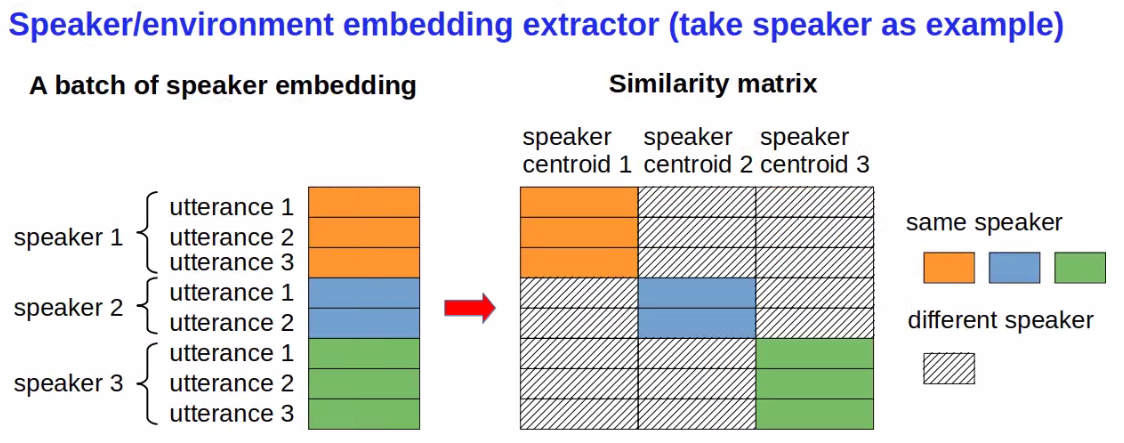}
  \caption{Training detail of speaker/environment embedding extractor.}
  \label{fig:speaker_embedding_extractor_training_detail}
  \vspace{-1.5em}
\end{figure}

\noindent \textit{Softmax loss for speaker embedding extractor:}
\begin{center}
$L(E_s^{ij})=-log \frac{exp(S_s^{ij,i})}{\sum\limits_{k=1}^{s} exp(S_s^{ij,k})}, \quad L(E_s)=\sum\limits_{\substack{1 \leq i \leq s \\1 \leq j \leq u}} L(E_s^{ij})$
\end{center}
$w$ and $b$ are parameters for training stability, where $w>0$. They are similar to the ones in \cite{wan2018generalized}. The speaker embedding extractor is trained to minimize the loss $L(E_s)$. The training detail is explained in Figure \ref{fig:speaker_embedding_extractor_training_detail}.

The Voxceleb1 dataset \cite{szoke2019building} is used to train the speaker embedding extractor. This dataset contains speech data from $1,251$ speakers. It particularly fits the purpose of modeling speaker-dependent factors, as it contains speech data of same speakers recorded under different acoustic environments.

\vspace{-1em}
\subsection{Environment embedding extractor}
\label{sec:environment_embedding_extractor}
The environment embedding extractor derives an environment embedding from a speech utterance. In the present study we assume that environment-specific information can be represented by the measured room impulse response (RIR). To train the extractor module, we purposely construct a speech dataset such that each type of environment is associated with multiple speakers. We use the clean speech utterances from 247 speakers in the ``train-clean-100'' subset of LibriTTS dataset \cite{zen2019libritts}. 2,325 kinds of measured RIRs from the BUT ReverbDB dataset \cite{szoke2019building} are used to represent different environments. For each environment condition, reverberant speech from multiple speakers are generated by convolving the clean speech with the respective RIR. The environment embedding extractor is trained in a similar way to the speaker embedding extractor, i.e., to force embeddings from the same environment to be close to each other and enlarge the distance among those from different environments. In each training batch, we select $e$ environments and sample $u$ utterances from each environment. Each utterance-level environment embedding $E_e^{ij}$, where $1 \leq i \leq e$ and $1 \leq j \leq u$, is compared to the centroids of all environments to obtain the similarity matrix and softmax loss, as illustrated in Figure \ref{fig:speaker_embedding_extractor_training_detail}.

\noindent \textit{Centroid:}
\begin{center}
$C_e^{i}=\frac{1}{u} \sum\limits_{1\leq m\leq u} E_e^{im}, \quad 
C_e^{i,(-j)}=\frac{1}{u-1} \sum\limits_{\substack{1\leq m\leq u, m \ne j}} E_e^{im}$
\end{center}

\noindent \textit{Similarity matrix of environment embeddings:}
$$ S_e^{ij,k}=\left\{
\begin{array}{rcl}
w*cos(E_e^{ij}, C_e^{i,(-j)})+b, & k=i\\
w*cos(E_e^{ij}, C_e^k)+b, & k\ne i 
\end{array}
\right.$$

\noindent \textit{Softmax loss for environment embedding extractor:}
\begin{center}
$L(E_e^{ij})=-log \frac{exp(S_e^{ij,i})}{\sum\limits_{k=1}^{e} exp(S_e^{ij,k})}, \quad L(E_e)=\sum\limits_{\substack{1 \leq i \leq e \\1 \leq j \leq u}} L(E_e^{ij})$
\end{center}
The environment embedding extractor is trained to minimize the loss $L(E_e)$.

\vspace{-1em}
\subsection{TTS module}
The TTS module is designed as a combination of the Tacotron2 \cite{shen2018natural} and the DurIAN \cite{yu2020durian} architectures. It is similar to the one proposed and evaluated in our previous work on voice cloning \cite{tan2021cuhk}, except that the embedding extractor modules replace the look-up tables.
The TTS module comprises an encoder, a duration predictor, a length regulator, a Prenet and a decoder. The duration predictor is used to determine phone-level duration from input text. The input text and phone-level duration are taken up by the encoder and the length regulator. A speaker embedding and an environment embedding are generated by the two extractors as described in Section \ref{sec:speaker_embedding_extractor} and \ref{sec:environment_embedding_extractor}. Both embeddings are broadcasted and concatenated with the encoder output to form the input to the decoder, which is similar to the method in \cite{jia2018transfer}. The Prenet and the decoder generate the mel-spectrogram of output speech, which is converted into speech waveform using the HiFi-GAN vocoder \cite{kong2020hifi}. The L2-norm loss between the generated spectrogram and the ground-truth one is minimized to train the TTS module. 

The loss for the overall proposed system is as follows:

\begin{center}
$L=L(E_s)+L(E_e)+L_{recon}(mel|E_s,E_e,text)$
\end{center}

\vspace{-1em}
\section{Experimental setup}

\subsection{Speech dataset}
\label{sec:dataset_construction}
The TTS module is trained with a specially constructed dataset in which the speaker factor and the environment factor are fully entangled. That is, each speaker is associated with a specific environment. Clean speech data of multiple speakers are obtained from the VCTK dataset \cite{yamagishi2019cstr}. Measured RIRs representing different environments are obtained from the REVERB challenge database \cite{kinoshita2013reverb} and the Aachen impulse response database \cite{jeub2009binaural}. 108 speakers and 108 environments are randomly selected to form 108 unique pairs of speaker-environment combination, and 100 utterances are sampled to represent each combination. It should be noted that, ``clean'' is regraded as one of the environments, for which clean speech is used directly in TTS training. For other environments, clean speech utterances from the respective speaker are convolved with the given RIR to generate training speech. $95$\% of the speaker-environment pairs are selected, and $95$\% utterances in these combinations are used as training data. The training data for TTS module do not have overlap with those for training the two embedding extractors, in terms of speaker identity and environment type.

\vspace{-1em}
\subsection{System configuration}
The audio data used a sampling rate of $22,050$ Hz. Short-time Fourier transform (STFT) is carried out with Hann window of 50 ms long and frame hop of 12.5 ms. Mel-spectrogram is computed from the magnitudes of STFT coefficients using 80-channel mel-scale filterbank from $0$ to $8,000$ Hz, followed by log dynamic range compression. The mel-spectrograms are used for the training of the speaker embedding extractor, environment embedding extractor and the TTS module. 

In the training of speaker (or environment) embedding extractor, each batch involves 64 speakers (or environments), and 10 utterances are sampled from the speech data of each speaker (or environment). From each utterance, a mel-spectrogram with 80-frame length is cropped for training. Thus total of 640 mel-spectrograms are included in each batch. Each of the embedding extractors contains three layers of 256-dimension unidirectional LSTM layers and one linear layer. The speaker (or environment) embedding is a 256-dimension vector. 

\vspace{-1em}
\subsection{Baseline system}
The baseline system consists of a speaker embedding extractor, an environment embedding extractor and a TTS module. These modules are trained jointly as describe with the same dataset in Section \ref{sec:dataset_construction}. The loss for the overall baseline system comprises three parts: speaker classification loss for speaker embedding extractor, environment classification loss for environment embedding extractor, and the speech reconstruction loss for the TTS module. The hyperparameter values used are the same as in the proposed system. The difference of the baseline system and proposed system is that, in baseline system, the speaker and environment embedding extractor is not trained with the objective mentioned in Section \ref{sec:speaker_embedding_extractor} and \ref{sec:environment_embedding_extractor}, but only two classification against speaker and environment are adopted for disentanglement supervision. 

The loss for the overall baseline system is as follows:

\begin{align} \nonumber
  L=&L_{cls}(E_s,speaker)+L_{cls}(E_e,environment)\\  \nonumber
  &+L_{recon}(mel|E_s,E_e, text)  \nonumber
\end{align}
\vspace{-1em}

\vspace{-1em}
\section{Results and discussion}

\subsection{Visualization of extracted embeddings}

\begin{figure}[h]
\centering
\subfigure[(a)]{\centering
\includegraphics[width=0.45\linewidth,height=0.45\linewidth, trim=50 50 50 50 ]{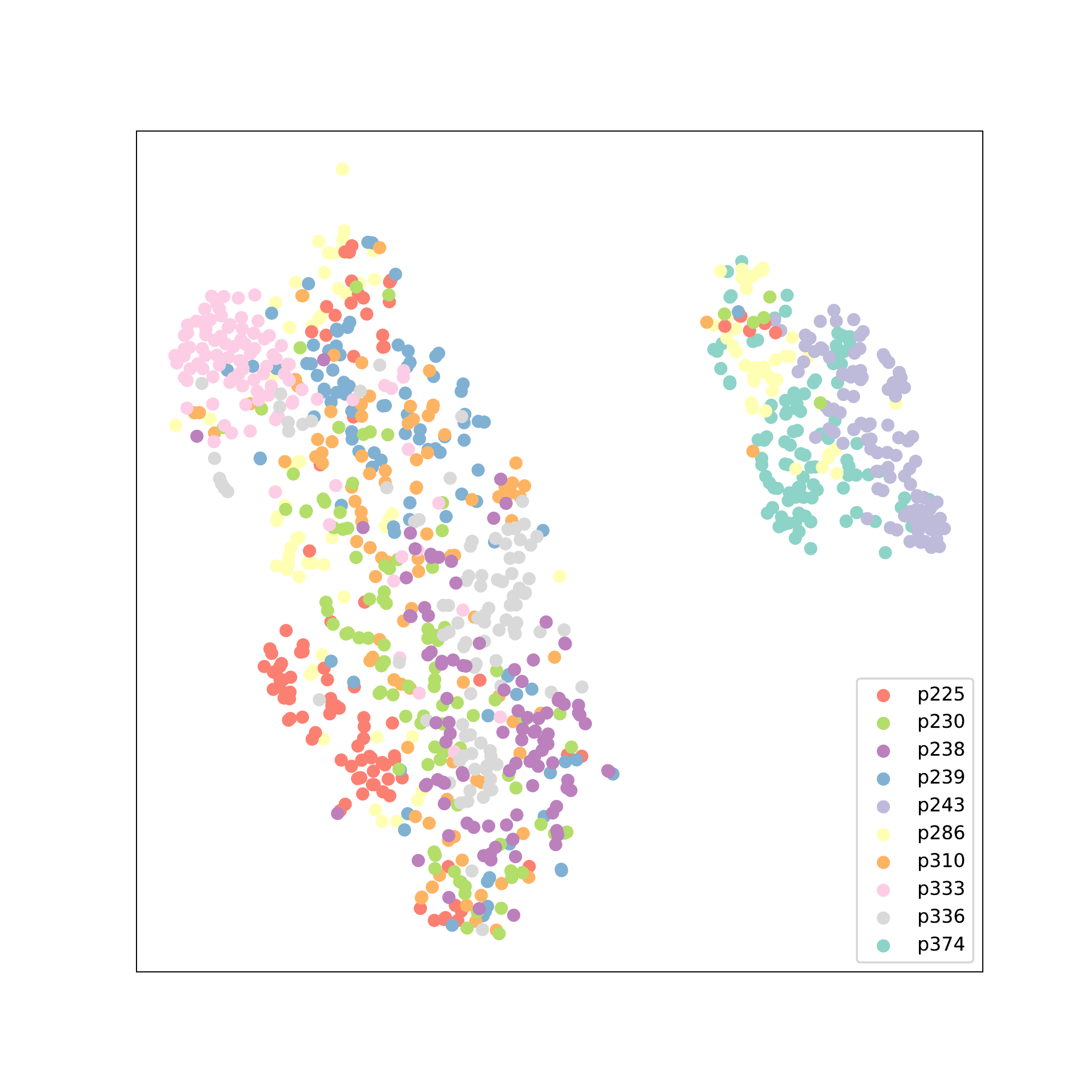}
}
\hspace{0.03\linewidth}
\subfigure[(b)]{\centering
\includegraphics[width=0.45\linewidth, height=0.45\linewidth, trim=50 50 50 50]{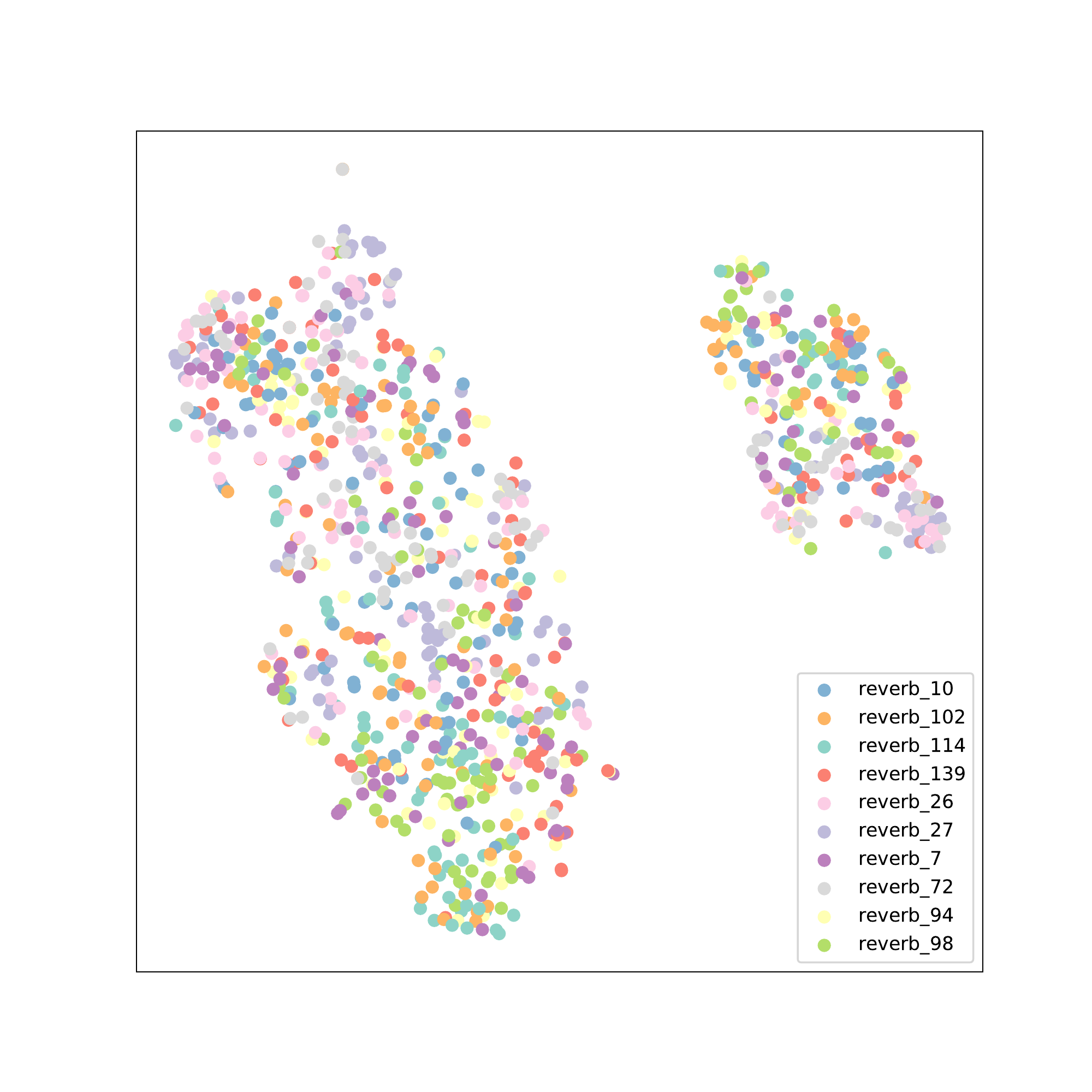}
}\\
\vspace{-0.05em}
\subfigure[(c)]{\centering
\includegraphics[width=0.45\linewidth, height=0.45\linewidth, trim=50 50 50 50]{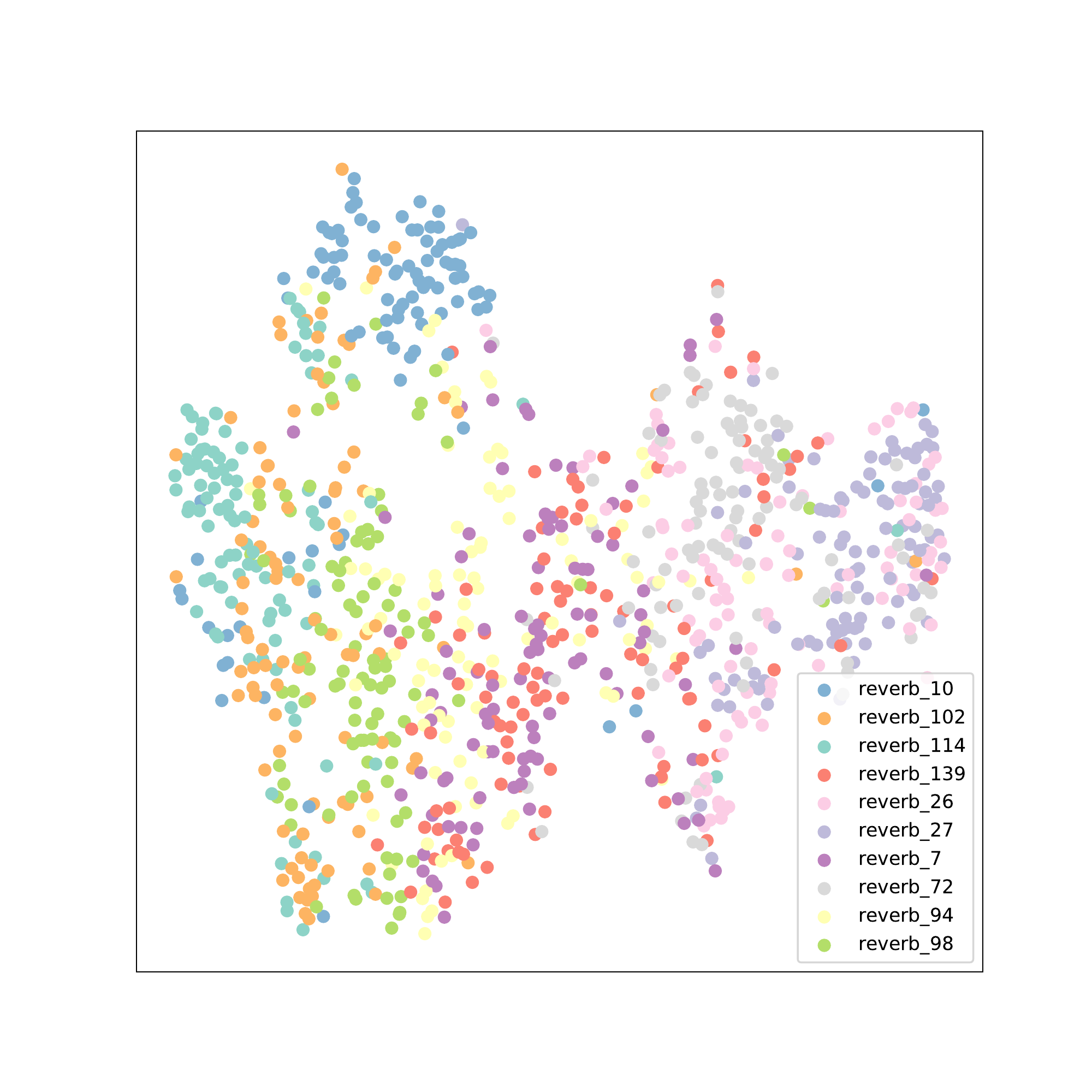}
}
\hspace{0.03\linewidth}
\subfigure[(d)]{\centering
\includegraphics[width=0.45\linewidth,height=0.45\linewidth, trim=50 50 50 50]{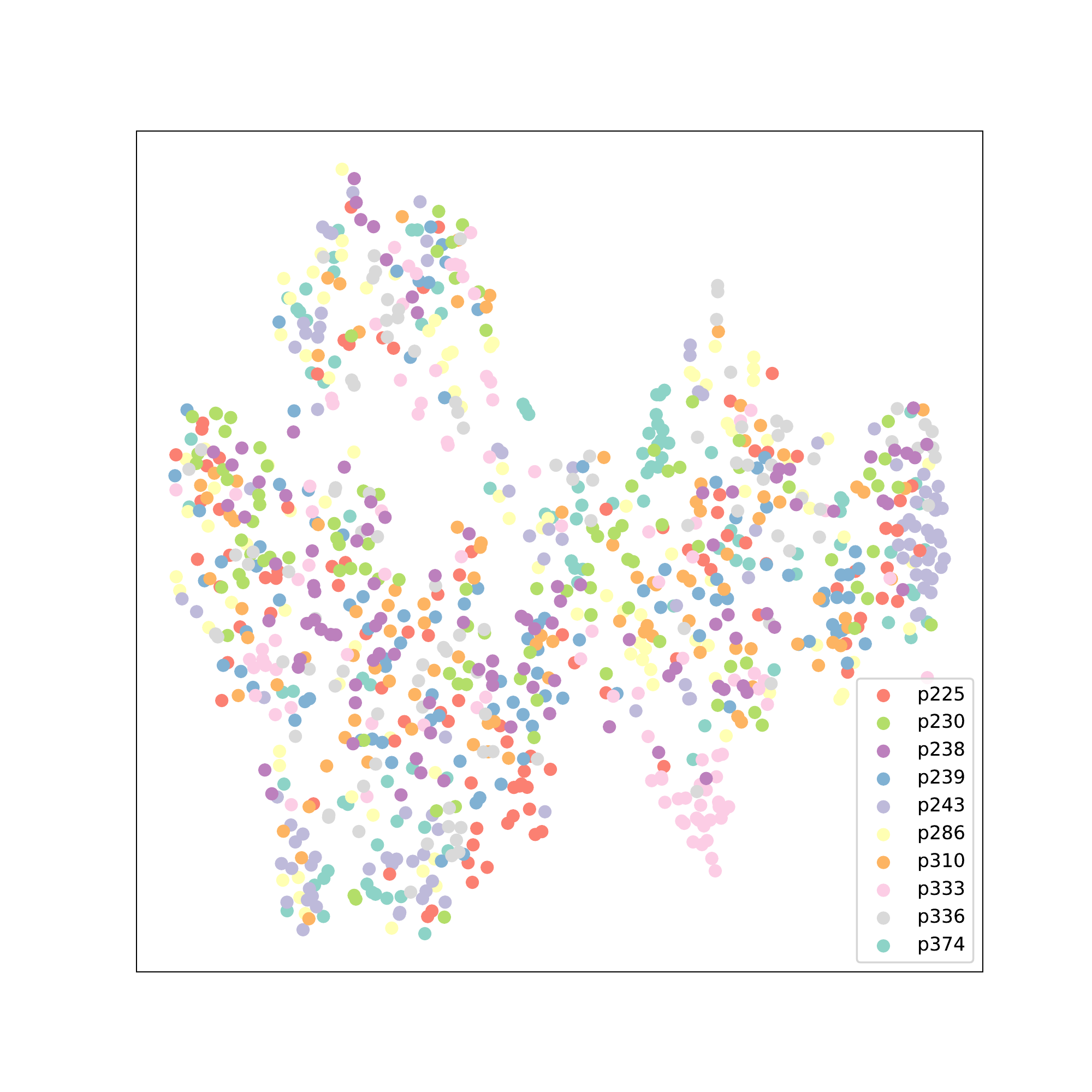}
}
\vspace{-1em}
\caption{The t-SNE visualization of speaker embedding and environment embedding. (a) speaker embedding with speaker label. (b) speake embedding with environment label. (c) environment embedding with environment label. (d) environment embedding with speaker label. }
\label{fig:tsne_embedding}
\vspace{-1em}
\end{figure}

To figure out whether the speaker embedding and environment embedding model and control corresponding factor of the speech, we first synthesize utterances conditioned on the speaker embedding and environment embedding that are extracted from two different utterances. Then we utilize the speaker embedding extractor and environment embedding extractor, mentioned in Section \ref{sec:speaker_embedding_extractor} and \ref{sec:environment_embedding_extractor}, to derive speaker embedding and environment embedding respectively from the synthesized utterance. T-SNE visualization \cite{van2008visualizing} is carried out on these two embedding along with the speaker and environment label. The result is shown in Figure \ref{fig:tsne_embedding}. It is noted from Figure \ref{fig:tsne_embedding} (a) and (b) that speaker embeddings from the same speakers tends to cluster together, and speaker embeddings concerning the same environment are mixed with each other. This suggests that the extracted speaker embeddings catpture speaker-related information and not environment-related one. From Figure \ref{fig:tsne_embedding} (c) and (d), it can be concluded that the extracted environment embeddings are able to capture the environment-related information but not speaker-related one in the speech. 

\renewcommand{\thesubfigure}{} 

\vspace{-1em}
\begin{table}[h]
\centering
\caption{Mel-cepstral distortion (MCD) between synthesized and ground-truth speech}
\label{tab:MCD}
\scalebox{0.9}{
\begin{tabular}{c|c|c|c|c}
\hline
\multirow{2}*{Speaker} & \multirow{2}*{Environment} & \multirow{2}*{Combination} & \multicolumn{2}{c}{MCD$\downarrow$}\\
\cline{4-5}
~ & ~ & ~ & Proposed & Baseline\\
\hline
seen & seen & seen  & 6.899 & \textbf{6.668} \\
\hline
seen & seen & unseen  & \textbf{7.312} & 7.605\\
\hline
unseen & unseen & unseen & \textbf{7.442}& 7.464\\
\hline
\end{tabular}
}
\vspace{-1em}
\end{table}

\begin{table}[h]
\centering
\caption{Embedding classification against speaker and environment of proposed and baseline system}
\label{tab:embedding classification}
\scalebox{0.9}{
\begin{tabular}{c|c|c|c|c|c}
\hline
\multirow{3}*{\makecell[c]{Combination}}  & \multirow{3}*{System}& \multicolumn{4}{c}{\makecell[c]{Classification accuracy$\uparrow$}} \\
\cline{3-6}
~& ~ & \multicolumn{2}{c|}{\makecell[c]{Speaker}} & \multicolumn{2}{c}{\makecell[c]{Environment}}\\
\cline{3-6}
 ~& ~ & Top-1 & Top-5 & Top-1 & Top-5 \\
\hline
\multirow{2}*{seen}  & Proposed & \textbf{51.0}\% & \textbf{71.2}\% & \textbf{48.4}\% & 66.0\% \\
\cline{2-6}
 ~  & Baseline & 43.8\% & 67.3\%  & 47.7\% & \textbf{72.5}\% \\
\hline
\multirow{2}*{unseen} & Proposed & \textbf{16.8\%} & \textbf{31.2\%} & \textbf{18.4\%} & \textbf{43.2\%}  \\
\cline{2-6}
~  & Baseline & 4.8\% & 13.6\% & 11.2\%  & 38.4\%  \\
\hline
\end{tabular}
}
\vspace{-1em}
\end{table}

\begin{table}[!h]
\centering
\caption{MOS with 95\% confidence interval on speaker similarity and environment similarity between synthesized speech and reference speech}
\label{tab:subjective evaluation}
\scalebox{0.9}{
\begin{tabular}{c|c|c|c}
\hline
\multicolumn{2}{c|}{MOS$\uparrow$} & Proposed & Baseline \\
\hline
\multirow{2}*{\makecell[c]{Seen \\combination}}& \makecell[c]{speaker similarity} & \makecell[c]{\textbf{3.88}$\pm$\textbf{0.09}} & \makecell[c]{3.76$\pm$0.10}\\
\cline{2-4}
~& \makecell[c]{environment similarity} & \makecell[c]{\textbf{3.74}$\pm$\textbf{0.10}} & \makecell[c]{3.47$\pm$0.11}\\
\hline
\multirow{2}*{\makecell[c]{Unseen\\ combination}}&\makecell[c]{speaker similarity} & \makecell[c]{\textbf{3.18}$\pm$\textbf{0.10}} &  \makecell[c]{1.98$\pm$0.11}\\
\cline{2-4}
~& \makecell[c]{environment similarity} & \makecell[c]{\textbf{3.25}$\pm$\textbf{0.10}} & \makecell[c]{2.33$\pm$0.11}\\
\hline
\end{tabular}
}
\vspace{-1em}
\end{table}

\subsection{Objective evaluation}
\vspace{-0.5em}
\subsubsection{Mel-cepstrum distortion}
\label{sec:MCD}
Objective evaluation on synthesized speech utterances is carried out using the metric of mel-cepstral distortion (MCD). The following three cases are investigated. In all cases, the ground-truth speech is obtained by convolving the clean speech from the respective speaker with the selected RIR.

\begin{enumerate}[leftmargin=*]
\item The synthesized speech is generated with speaker-environment combinations covered in the training dataset (i.e., ``seen combinations''). The MCD result can reflect the system's general performance in speech generation;
\item The synthesized speech is generated with new speaker-environment combinations that are not covered by the training dataset (i.e., ``unseen combinations''). More precisely, both the speaker identity and the environment type are included in the training data but their combination is not. The MCD result can reflect the system's efficacy of disentanglement of speaker and environment factors;
\item The synthesized speech is generated with both the speaker identity and the environment type that are not included in system training (i.e., ``unseen speaker'' and ``unseen environment''). The MCD result can reflect the generalization capacity of the systems.
\end{enumerate}

The results of objective evaluation are given in Table \ref{tab:MCD}. The proposed system is not as good as the baseline system in generating speech on the seen speaker-environment combinations. However, for unseen speaker-environment combination, the proposed system can attain significant lower MCD than the baseline system, indicating that it is able to disentangle the speaker and environment factors and model them independently. The proposed system performs slightly better than the baseline system for unseen speaker and unseen environment, indicating that the proposed system has slightly better generalization capacity for speaker and environment.

\subsubsection{Classification of embeddings}
Two classifiers are trained on speaker embeddings and environment embeddings respectively. The embeddings extracted from natural speech are used as the input for the training of classifiers, while the ground-truth speaker or environment labels are used as training output for classification.

With the trained classifiers, two kinds of synthesized speech are examined. The first one is the same as Case 1 in Section \ref{sec:MCD}, i.e., with seen speaker-environment combinations. The second one is the same as Case 2, i.e., unseen combination of seen speaker and seen environment. The trained speaker and environment extractors are used to derive the embeddings from synthesized utterances. The embedddings are used as the input of the corresponding classifier to predict the speaker identity or environment type. The classification results are shown as in Table \ref{tab:embedding classification}. For the seen combinations of speaker and environment, the proposed system exhibit better accuracy on embedding classification against speaker and environment than the baseline system. For unseen combinations of speaker and environment, the proposed system performs significantly better than the baseline system and demonstrates much better disentanglement capacity of speaker and environment factor.

\vspace{-0.5em}
\subsection{Subjective evaluation}
Subjective evaluation is performed on two scenarios. The first scenario is speech generation with seen speaker-environment combinations. In this scenario, a natural reference utterance is used for extracting speaker and environment embeddings, and both embeddings are used in speech generation. Two versions of synthesized utterances are obtained from the proposed and baseline systems respectively. Participants of the evaluation are required to listen to the audio samples and rate the speaker similarity and the environment similarity between each synthesized utterance and the reference one. 

The second scenario is speech generation with new combination of seen speaker and seen environment. Two reference utterances are involved. One is used for extracting speaker embedding and the other one for extracting environment embedding. The two embeddings are then used to produce two synthesized utterances with the proposed and the baseline systems. The listeners are required to rate the speaker or environment similarity between each synthesized utterance and the respective reference speech. 

In both scenarios, the ratings on speaker and environment similarity range from 1 (``completely different'') to 5 (``exactly the same''). 15 listeners participated in the listening test and each of them rated $32$ samples. The result is shown as in Table \ref{tab:subjective evaluation}. It can be seen that the proposed system performs better than the baseline system, in terms of both speaker and environment similarity in both scenarios. The proposed system demonstrates better disentanglement capacity and generation performance.

\vspace{-0.5em}
\section{Conclusion}
In this paper, we propose an environment-aware TTS system, which is capable of generating speech that carries designated speaker timbre and environment attribute. Characterization and disentanglement of speaker and environment factors in speech are carried out in the system design. Both objective and subjective evaluation results have shown that our proposed system is not only able to model the speaker and environment in speech respectively, but also performs better than baseline system in terms of the speaker and environment similarity of the generated speech. Our system can be further used in text-based speech editing, robust voice cloning and speech de-reverberation.

\vspace{-0.5em}
\section{Acknowledgements}
This research is partially supported by a Knowledge Transfer Project Fund (Ref: KPF20QEP26) from the Chinese University of Hong Kong.

\vfill\pagebreak

\bibliographystyle{IEEEtran}

\bibliography{mybib}

\end{document}